\documentstyle[12pt,epsfig]{article}
\newcommand{\be}{\begin{equation}}
\newcommand{\ee}{\end{equation}}
\begin{document}  
\topmargin 0pt
\oddsidemargin=-0.4truecm
\evensidemargin=-0.4truecm
\renewcommand{\thefootnote}{\fnsymbol{footnote}}
%\newpage
\setcounter{page}{0}
\begin{titlepage}   
\vspace*{-2.0cm}  
%%%\vspace*{-1.0cm}
\vspace*{0.1cm}
\begin{center}
{\Large \bf Neutrino Oscillations, Solar Antineutrinos and the Solar Magnetic Fields}\\
\vspace{0.6cm}

{\large 
S. Dev and Sanjeev Kumar \\
%EndAName
Department of Physics, Himachal Pradesh University, Shimla-5, India.}
\end{center}

\vglue 0.6truecm
\begin{abstract}

Even after the confirmation of the large mixing angle (LMA) solution of the
solar neutrino problem, the scope for resonant spin- flavor precession
(RSFP) transitions as a subdominant effect still exists. In this work,we
have considered suitably suppressed RSFP transitions in addition to the
dominant LMA flavor transitions and translated the bounds on the
antineutrino flux into the bounds on the product of neutrino magnetic moment
and solar magnetic field. The low and intermediate energy neutrinos have
been included in the analysis by obtaining indirect bounds on the
corresponding antineutrino fluxes for these components. It is assumed that
the missing beryllium neutrinos are being converted into muon antineutrinos
via RSFP transitions in the Sun.
\end{abstract}

\end{titlepage}   
\renewcommand{\thefootnote}{\arabic{footnote}}

\section{Introduction}

The neutral current measurements at SNO \cite{1} have confirmed the
oscillations of solar neutrinos. After the first evidence of antineutrino
disappearance in a beam of electron antineutrinos reported by KamLAND
experiment \cite{2}, all other explanations of the solar neutrino problem
are either completely discarded or may just be subdominant effects. After
these pioneering experiments, there is hardly any scope for doubt about the
physical reality of neutrino mass and their, consequent, oscillations. Once
neutrino mass is observed, neutrino magnetic moments are an inevitable
consequence in the Standard model and beyond.

KamLAND is the first experiment to explore the neutrino parameter space
relevant to solar neutrinos with a beam of terrestrial neutrinos. KamLAND
has, convincingly, demonstrated the existence of neutrino oscillations with
parameters confined to the LMA region. Apart from the measurement of reactor
antineutrino flux, KamLAND is sesitive to any boron antineutrinos produced
from solar boron neutrinos as a result of spin flavor oscillations. The flux
of solar antineutrinos is expected to be considerable, in case, the
neutrinos possess Majorana character and, in addition, the neutrino magnetic
moment is high enough. Thus, the observation of a small solar antineutrino
flux is likely to establish the Majorana character of the neutrino and the
existence of physics beyond the Standard Model. An important signature of
RSFP is an observable flux of electron antineutrinos from the Sun if
neutrinos, in addition to transition magnetic moments, have a sizable flavor
mixing. The flux of solar electronic antineutrinos can, in principle, be
detected at BOREXINO and SNO. If both transition magnetic moments and flavor
mixing of massive neutrinos are present, the combined action of RSFP and the
flavor oscillations may lead to an observable flux of electron neutrinos.
The existing bounds on the solar antineutrino flux from KamLAND are fairly
stringent \cite{3}.

It may be worthwhile to emphasize again that the observation of solar
electronic antineutrino flux will be a unique signature for a Majorana
neutrino transition magnetic moment and the resonant spin-flavor precession
inside the Sun. Therefore,the results of the electronic antineutrino flux
measurements by BOREXINO will be eagerly awaited since the sensitivity of
BOREXINO to the electronic antineutrinos is expected to be about 37 times
larger than that to the electronic neutrinos. The detection of solar
antineutrinos is one of the major goals of several existing and forthcoming
solar neutrino experiments.

The KamLAND bounds on the boron antineutrino flux are available now \cite{3}
and the beryllium antineutrinos will be observable in BOREXINO. SNO has
already indicated that the solar electron neutrinos are being converted into
active neutrinos by observing the neutral current processes \cite{2}.
Therefore, only a small fraction of electron neutrinos is allowed to be
converted into the muon antineutrinos via the RSFP mechanism within the
solar interior. These muon neutrinos can, then, oscillate into the electron
antineutrinos while travelling from the solar surface to the earth via
vacuum transitions with a probability of about half. The bounds on the
antineutrino flux have been used to obtain the bounds on $\mu B$ for the
case of high energy boron neutrinos \cite{3}. After KamLAND results and
earlier SNO neutral current measurements, there is hardly any doubt about
the fact that the neutrino oscillations with LMA parameters explain the
solar neutrino deficit and that the electron neutrinos are being,
dominantly, converted into muon neutrinos. Only a small fraction is allowed
to be converted into antineutrinos via RSFP mechanism.

In this work, we consider the solar antineutrino production in the RSFP
scenario\cite{4} in coexistence with the LMA MSW scenario and reexamine the
earlier results for boron neutrinos. We also extend our analysis to include
the low and intermediate energy neutrinos which have not been considered in
the earlier analyses \cite{3}. In order to obtain bounds on low and
intermediate energy antineutrino fluxes, we compare the model independent
values of the survival probabilities with their values in the pure LMA case
assuming that the difference between the two is due to the RSFP driven $%
v_{e}\rightarrow \overline{v_{\mu }}$ transitions. It is noted that
beryllium neutrino flux is smaller than the LMA expectations. This could
explain the low Ar-production rate in the Homestake experiment sought to be
explained by Smirnov \cite{5} by invoking the existence of a light sterile
neutrino. In this work, we surmise that the missing beryllium neutrinos are
being converted into muon antineutrinos via RSFP transitions within the Sun.
The muon antineutrinos, thus produced, undergo vacuum oscillations into the
electron antineutrinos in their flight to the earth. This antineutrino flux
can be detected by BOREXINO. Thus, the hypothesis advanced in the present
work can be directly confirmed or discarded by BOREXINO in contrast to the
hypothesis put forth by Smirnov \cite{5} where the missing beryllium
neutrinos are assumed to be oscillating into sterile neutrinos. In this
scenario, the neutrino transition magnetic moments and solar magnetic fields
required to account for the $v_{e}\rightarrow \overline{v_{\mu }}$
conversion of missing beryllium neutrino flux are well within their known
limits . In order to avoid an anomalously large RSFP conversion of boron
neutrinos, the magnetic field has to undergo a decrease towards the edge of
the core. A suitably decreasing magnetic field profile also explains the
absence of the rise up of neutrino energy spectrum at SuperKamiokande. This
rise up can be as large as 10 percent in the pure LMA scenario \cite{5}.

\section{Solar Magnetic Fields and Antineutrino Production}

In our model of antineutrino production, the $\nu _{e}\rightarrow \nu _{\mu
} $ flavor transitions are described by the LMA scenario while the $\nu
_{e}\rightarrow \overline{v}_{\mu }$ transitions are described by the RSFP
scenario where both types of transitions are assumed to be occurring
independently of each other. This approximation is well justified for the
LMA oscillation parameters for which the RSFP resonance width is too small
to interfere with the LMA flavor transitions \cite{6}.

The RSFP resonance condition \cite{6} 
\begin{equation}
\frac{5G_{F}N_{e}}{3\sqrt{2}}=\frac{\Delta m^{2}\cos 2\theta }{2E}
\end{equation}
implies that for the neutrinos of energy E , the resonance will occur at the
point 
\begin{equation}
x=0.09\ln \left( \frac{E}{0.45}\right)
\end{equation}
where $\Delta m^{2}=7.1\times 10^{-5}eV^{2},\sin ^{2}2\theta =0.9$, and we
have considered the standard model density profile \cite{7}. Here $x=\frac{r%
}{R_{S}}$ and the energy E is in the units of MeV. Boron neutrinos have
average energy 6.7 MeV. The production region of the boron neutrinos extends
up to x=0.15 \cite{7} . Thus, in the course of their propagation, the boron
neutrinos will get the resonance density at x=0.24. The resonance point for
the beryllium neutrinos $(E=0.861MeV)$ will be at x=0.06 which is near the
point of maximal production \cite{7}. Thus, the production and the resonance
regions may coincide for the $^{7}$Be (0.861 MeV) neutrinos for certain
values of LMA parameters . This makes the LMA transition probability for
beryllium neutrinos about half. However, the adiabatic RSFP transition
probability for beryllium neutrinos is very small for values of $\mu B$
within the range constrained by various laboratory and astrophysical bounds.
Therefore, the non-adiabatic effects in the resonance region will be
important. As the energies of the low energy pp neutrinos are below 0.45
MeV, they will not get the resonance densities anywhere.

The production region for the neutrinos is an extended one. For the pp
neutrinos the maximal production occurs at the point x=0.1 whereas for the
boron and beryllium neutrinos, the maximal production occurs at about x=0.05 
\cite{6}The mixing angle in matter in the presence of a magnetic field is
given by \cite{5} 
\begin{equation}
\tan 2\widetilde{\theta }_{m}=-\frac{s_{R}}{s_{\Delta }}
\end{equation}
where 
\begin{equation}
s_{R}=2\mu B,s_{\Delta }=\frac{5G_{F}N_{e}}{3\sqrt{2}}-\frac{\Delta
m^{2}\cos 2\theta }{2E}
\end{equation}

The RSFP transition probability is given by 
\begin{equation}
P(\nu _{e}\rightarrow \overline{\nu }_{\mu })=\frac{1}{2}-\left( \frac{1}{2}%
-P_{c}\right) \cos 2\widetilde{\theta }\cos 2\widetilde{\theta }_{edge}
\end{equation}
where $\cos 2\widetilde{\theta }$ is to be evaluated at the production point
whereas $\cos 2\widetilde{\theta }_{edge}$ is to be evaluated at the edge
and is approximately one. The crossing probability given by 
\begin{equation}
P_{c}=\exp \left( -\frac{2s_{R}^{2}E}{\Delta m^{2}\cos 2\theta }%
0.09R_{s}\right)
\end{equation}
is to be evaluated at the respective resonances \cite{5}. For LMA
parameters, it simplifies to $P_{c}=\exp \left( -4440.7s_{R}^{2}E\right) $
where $s_{R}$ is in the units of $10^{-11}eV$ and E is in the units
of MeV. For beryllium neutrinos, we have $s_{\Delta }=0.12\times 10^{-11}eV$
.%
 For the boron and pp neutrinos, we have $s_{\Delta }=1.28\times
10^{-11}eV,-3.42\times 10^{-11}eV$ at the respective production points for
their respective average energies.

The bounds on $\mu $ are available from the various sources \cite{8}. The
reactor bounds on the neutrino magnetic moment are $\mu \leq 10^{-10}\mu
_{B} $. The bounds on $B$ \cite{9} are less certain and little is known
about its magnitude and spatial variation, especially in the deep solar
interior. We can safely assume that $B\leq 1000T$ in the solar core. For $%
B=10000T$, the SFP adiabatic transition probability is 0.64 for pp
neutrinos. A considerable transformation into antineutrinos is,thus,
allowed, even in the adiabatic approximation. Hence, such a large field is
directly ruled out by virtue of the reactor bounds on the neutrino
transition moments. This lowers considerably the Chandershekhar bound on B 
\cite{10}.

We, therefore, take $\mu =10^{-10}\mu _{B},B=1000T$ as an upper bound, which
implies $s_{R}=1.16\times 10^{-11}eV$. There will be considerably large
adiabatic transitions for this value of $s_{R}$. The values of cos2$\theta $
for pp , beryllium and boron neutrinos are 0.96, -0.12 and -0.78
respectively. Thus, the transition probability is 0.02 for pp neutrinos in
the adiabatic approximation which is justified for pp neutrinos because they
do not undergo resonance. For Be and B neutrinos,the crossing probability is
not zero and will be the only effect for smaller values of $\mu B$. For pp
neutrinos, the transition probability (adiabatic) will approach zero as $\mu
B$ decreases. If $2\mu B$ is one order of magnitude below the above upper
bound, {\it i.e.} $2\mu B\leq 10^{-12}eV($e.g.$\mu \leq 10^{-12}\mu
_{B},B\leq 1000T)$, the transition probability is $\leq 2.2\times 10^{-6}$
for the pp neutrinos. For the beryllium and boron neutrinos, $\cos 2\theta
\sim -1$ so that the adiabatic transition probability will be zero for them.
Therefore, the transition of pp neutrinos into antineutrinos is not expected
at all and the transformation of beryllium and boron neutrinos is highly
non-adiabatic (adiabatic transition probability being zero). Hence, for the
intermediate and high energy neutrinos, we have 
\begin{equation}
P(\nu _{e}\rightarrow \overline{\nu }_{\mu })=1-P_{c}
\end{equation}
for $\mu \leq 10^{-12}\mu _{B},B\leq 1000T$.

The bounds on antineutrino flux produced from the $^{7}$Be neutrinos
will be provided by BOREXINO. From these bounds, one can constrain $\mu B$
within the core. Conversely, assuming suitable values for $\mu B$, one can
predict the beryllium electron antineutrino probability and, hence, the
beryllium electron antineutrino flux.

The probability for the electron antineutrinos to appear at the earth is
given by 
\begin{equation}
P(\nu _{e}\rightarrow \overline{\nu }_{e})=P_{RSFP}(\nu _{e}\rightarrow 
\overline{\nu }_{\mu })P_{VAC}(\overline{\nu }_{\mu }\rightarrow \overline{%
\nu }_{e})
\end{equation}
where, $P_{VAC}(\overline{\nu }_{\mu }\rightarrow \overline{\nu }_{e})=\frac{%
1}{2}\sin ^{2}2\theta $.

Alternative models of antineutrino production studied in the literature \cite
{3} either consider a highly chaotic magnetic field in the radiation zone of
the Sun or use perturbation calculations [valid only for small $\mu B$ ] to
obtain the RSFP transition probability. In perturbative calculations, the
transition probability depends only on the magnetic field in the production
region. In both categories of models, the transition probabilities for
beryllium and boron neutrinos can not be much different. Thus, these models
can not explain a large transition probability for beryllium neutrinos into
antineutrinos while still keeping the boron antineutrino production at a
phenomenologically acceptable level, in contrast to the model considered in
the present work.

The model independent survival probabilities for the low, intermediate and
high energy neutrinos have been calculated from the solar neutrino data by
Barger {\it et al} \cite{11}. The survival probability for high energy
neutrinos can be obtained from SNO NC and CC measurements alone the value
for which in the pure LMA scenario is $\sin ^{2}\theta =0.34$. The agreement
between the two values is the first step towards the solution of the solar
neutrino problem. The experimental values, however, do not agree with the
survival probabilities in the pure LMA scenario for low and intermediate
energy neutrinos. The survival probability for the beryllium neutrinos is
expected to be nearly half in the pure LMA scenario compared to the value of
0.3 obtained from the model independent analysis of the solar neutrino data 
\cite{11} poinitng towards a significant portion of this component to be
missing \cite{12}. The two values do not agree even if the uncertainties in
the experiments and the SSM fluxes are taken into consideration. Similarly,
the LMA survival probability for pp neutrinos is the same as the vacuum
survival probability $\left( 1-\frac{1}{2}\sin ^{2}2\theta \right) =0.55$
whereas the experimental value is about 0.8 \cite{11}. Thus, the observed pp
neutrino flux is larger than the LMA expectation. Therefore, there is no
scope for antineutrino production in the low energy range.

The missing beryllium neutrino flux can account for the low Ar-production
rate at Homestake sought to be explained by Smirnov \cite{5} by invoking
mixing with a sterile component. It is clear that a portion of the beryllium
neutrinos is undergoing other subdominant transitions in addition to the
dominant LMA flavor oscillations. These transitions have to be effective
only at the intermediate energies. An attractive candidate for these are the
RSFP transitions, though, the transitions into a light sterile neutrino can
also explain this anomaly \cite{5}. The attractiveness of the RSFP
transitions of missing beryllium neutrinos into antineutrinos lies in the
possibility of their direct detection at BOREXINO. If the event rate at
BOREXINO is found to be smaller than $R_{BOREXINO}\simeq 0.5$ (pure LMA
value), it would confirm that a portion of the beryllium neutrinos are
missing. In addition, if BOREXINO,also, observes an antineutrino flux of
about 0.1 of the SSM beryllium neutrino flux, it would be confirmed that the
missing beryllium neutrinos are undergoing RSFP transitions. Otherwise, the
possiblity of oscillation into the sterile neutrinos has to be taken
seriously.

Similarly, the RSFP transitions of boron neutrinos will result in values of
the fluxes smaller than the LMA expectations. In the pure LMA scenario, the
flux of boron neutrinos is expected to increase slightly as the energy
decreases. This rise up in the neutrino energy spectrum can be as large as
10 percent \cite{5}. The possibility of RSFP transitions can result in a
decrease in the neutrino flux which can explain the lack of the rise up of
the boron neutrino spectrum at low energies.

~From equations (5), (6 ), and (7), it follows that to account for the
missing beryllium neutrinos via RSFP tarnsitions, we require $2\mu B\sim
7.6\times 10^{-14}eV,(e.g.\mu \sim 10^{-12}\mu _{B},B\sim 328T)$ deep inside
the core ($x\sim 0.05$). To explain the upper limit on the boron
antineutrino appearance probability (about 0.0034) obtained at KamLAND \cite
{3}, we must have $2\mu B\sim 3.18\times 10^{-15}eV,(e.g.\mu \sim
10^{-12}\mu _{B},B\sim 13.7T)$ towards the edge of the core ($x\sim 0.25$).
The recent direct bounds on the solar magnetic field at the bottom of
radiation zone are: $B=300-700T$ \cite{13} which are independent of the
values of the neutrino magnetic moment. If these bounds are to be taken
seriously, we must take $\mu \sim (3.9-9.2)\times 10^{-14}\mu _{B}$ to have
the required value of $2\mu B$ at $x\sim 0.25$. Thus, a magnetic field of
the order of about 10000T is required deep within the core.

~From the equations (5),(6 ) and (7), and the rise up in the pure LMA
survival probability, one can calculate the solar magnetic field profile. If
P$_{LMA}$(E) is the energy dependent survival probability in the LMA
scenario, we must have 
\begin{equation}
P_{LMA}=1-P_{c}+\sin ^{2}\theta
\end{equation}
in order to have no rise up in the boron neutrino spectrum at
SuperKamiokande where the effects of flux-averaging over the production
region have been neglected. The pure LMA survival probability \cite{14} as a
function of energy is given by 
\begin{equation}
P_{LMA}(E)=\frac{1}{2}+\frac{1}{2}\cos 2\theta \cos 2\theta _{m}
\end{equation}
where 
\[
\tan 2\theta _{m}=\frac{\sin 2\theta }{\cos 2\theta -2\sqrt{2}
G_{F}N_{e}E/\Delta m^{2}}
\]
The LMA survival probability $P_{LMA}(E)$ has been ploted in Fig.1.as a
function of energy E ( in the units of MeV). The rise up in the survival
probability for boron neutrinos at low energies is about 4 percent. Also,
for the average value of energy (6.7 MeV), $P_{LMA}=0.3480$ while $\sin
^{2}\theta =0.3418$. This implies an electron antineutrino appearance
probability of 0.0031 in fairly close agreement with the KamLAND bound \cite
{3}.

The magnetic field profile in the resonance region of boron neutrinos
extending from 0.2 to 0.3 can be obtained, directly, from equation (9). In
Fig.2., the magnetic field B( in the units of Tesla) has been plotted
against the distance from the center of the Sun in the units of the solar
radius, assuming $\mu \sim 10^{-13}\mu _{B}$.

\section{Conclusions}

To conclude, we considered the solar antineutrino production via RSFP
transitions in coexistence with the dominant LMA MSW transitions to explain
the KamLAND bounds on solar antineutrinos. The bounds on the antineutrino
flux have been translated into the bounds on solar magnetic fields. Recent
studies \cite{3} on solar antineutrinos are limited to boron antineutrinos
for which direct bounds are available from KamLAND. However, we have incuded
low and intermediate energy neutrinos in this analysis. Indirect bounds on
low and intermediate energy antineutrino fluxes have been obtained by
comparing the model independent values of the survival probabilities
obtained from the solar neutrino data with their values in the pure LMA
case. It is noted that beryllium neutrino flux is smaller than the LMA
expectations implying that the beryllium neutrinos are also undergoing
transitions other than the flavor oscillations at a subdominant level. A
natural candidate for these transitions are the spin-flavor transitions
which can result in lower event rate at BOREXINO as compared to the pure LMA
expectations. Whether this is actually happening can,directly, be verified
by BOREXINO. The neutrino transition moments and solar magnetic fields
required to account for the missing beryllium neutrino flux are well within
the known limits on these quantities. Moreover, to avoid an anomalously
large RSFP conversion of boron neutrinos, the magnetic field has to undergo
a decrease towards the edge of the core. In addition, a suitably decreasing
magnetic field profile can also explain the absence of the rise up of
neutrino energy spectrum at SuperKamiokande.

\section*{Acknowledgments}

One of the authors (SK) gratefully acknowledges the financial support
provided by the Council for Scientific and Industrial Research(CSIR),
Government of India.

\newpage
\begin{figure}[h]
\setlength{\unitlength}{1cm}
\begin{center}
%\hspace*{-1.8cm}
\hspace*{-1.6cm}
\epsfig{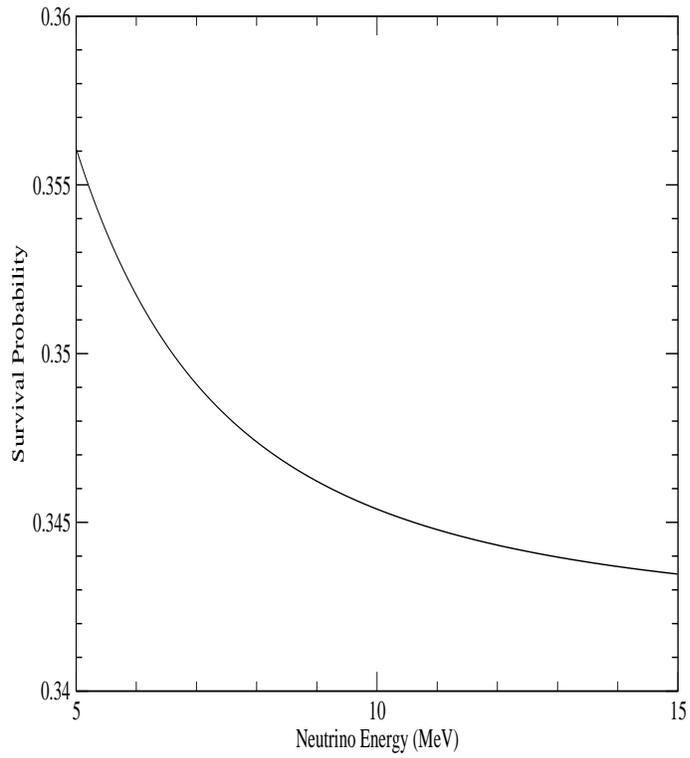}
\end{center}
\caption{ \it LMA survival Probability for $^8B$ neutrinos as a function of energy for $\sin ^{2}\theta~=~0.3418$.}  
\label{fig1}
\end{figure}

\begin{figure}[h]
\setlength{\unitlength}{1cm}
\begin{center}
%\hspace*{-1.8cm}
\hspace*{-1.6cm}
\epsfig{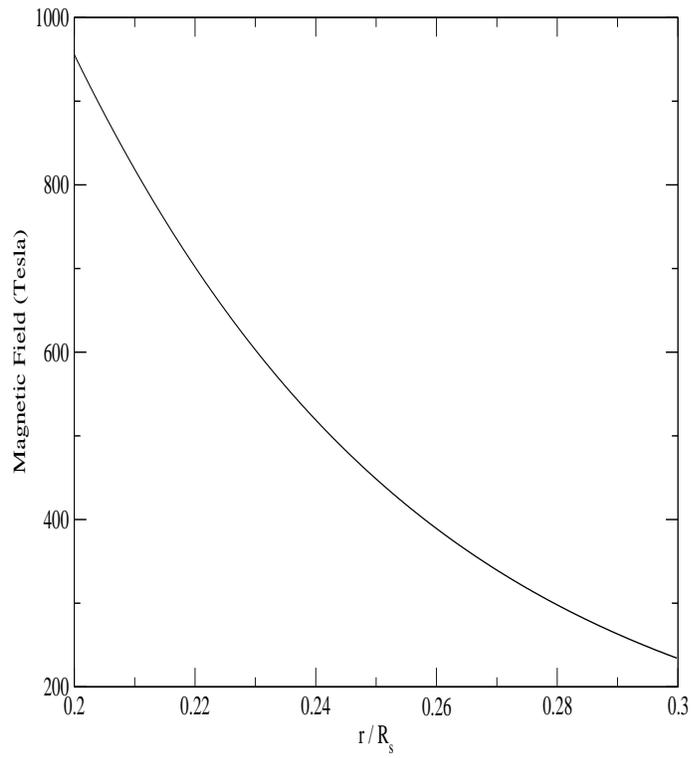}
\end{center}
\caption{ \it The magnetic field profile in the resonance region of $^8B$ neutrinos.}

\label{fig2}
\end{figure}

\end{document}